\documentclass[
reprint,
superscriptaddress,
amsmath,amssymb,
aps,
pra,
]{revtex4-2}

\usepackage{graphicx} % Required for inserting images
\usepackage{amsmath}  % improve math presentation
\usepackage{dsfont}
\usepackage{physics}
\usepackage{xcolor}
\usepackage{appendix}
\newcommand{\ben}{\begin{eqnarray}}
\newcommand{\een}{\end{eqnarray}}

\begin{document}

\title{Inter-qubit correlation dynamics driven by mutual interactions}

  \author{Aleksandra Kwiatkowska}

\affiliation{Institute of Theoretical Physics and Astrophysics, Faculty of Mathematics, Physics, and Informatics, University of Gdańsk, 80-308 Gdańsk, Poland}

  \author{Waldemar K{\l}obus}

\affiliation{Institute of Theoretical Physics and Astrophysics, Faculty of Mathematics, Physics, and Informatics, University of Gdańsk, 80-308 Gdańsk, Poland}

\begin{abstract}
A particularly useful tool for characterizing multi-qubit systems is the correlation tensor, providing an experimentally friendly and theoretically concise representation of quantum states.
In this work, we analyze the evolution of the correlation tensor elements of quantum systems composed of $\frac12$-spins, generated by mutual interactions and the influence of the external field. We focus on two-body interactions in the form of anisotropic Heisenberg as well as antisymmetric exchange interaction models. The evolution of the system is visualized in the form of a trajectory in a suitable correlation space, which, depending on the system's frequencies, exhibits periodic or quasiperiodic behavior. In the case of two $\frac12$-spins we study the stationary correlations for several classes of Hamiltonians, which allows a full characterization of the families of density matrices invariant under the evolution generated by the Hamiltonians. We discuss some common properties shared by the 2- and 3-qubit systems and show how a strong external field can play a stabilizing factor with respect to certain correlation characteristics. 
\end{abstract}

\maketitle

\section{Introduction}

Quantum correlations, whose most prominent manifestation is quantum entanglement, constitute a pivotal feature of quantum mechanics, enabling operations and effects unattainable with classical apparatus, thus playing a central role in quantum information protocols. Although quantum entanglement is a necessary resource, it is nevertheless the specific form of correlations between well-defined measurements that enables a particular information-processing task. For instance, the violation of Bell inequalities which underpins quantum advantage in, e.g., communication complexity problems, requires generating strong correlations between the outcomes of two pairs of predefined measurement settings \cite{RevModPhys.82.665,deVicente_2014,PhysRevResearch.4.L012002,RevModPhys.86.419,wfx7-1v2c}.
In this context, the central issue becomes the ability to generate the required correlations, while the precise form of the quantum state is of secondary relevance. From this point of view, it is essential to understand how the specific correlations emerge as a result of the interplay between local and global properties of the system or due to additional external factors.

In this work, we attempt to analyze how the specific correlations arise as a result of a natural interaction between 2-dimensional subsystems (qubits). While in the past much effort has been put into analyzing entanglement \cite{entmesmp,entmesmh,PhysRevA.83.012312,mpem,MA2024}, we take different approach, focusing on correlations defined by the expectation values of local (i.e. one-particle) observables. The set of all correlations measured in a predefined reference frame defines \textit{correlation tensor} \cite{PhysRevLett.100.140403,PhysRevA.84.062305,PhysRevA.84.062306}, which also provides a natural means to quantify the extent of Bell-inequality violation \cite{HORODECKI1995340} as in the aforementioned scenario.

In the case of systems composed of 2-dimensional subsystems, the elements of the correlation tensor pose as a natural extension of the Bloch vector components into a multipartite case. Unfortunately, for a system composed of $N$ qubits, the complete characterization involves $4^N-1$ parameters, the evolution of which cannot be aptly visualized in a low-dimensional space such as Bloch ball for a single qubit. One may, however, reduce the set of parameters by considering the lengths of Bloch vectors \cite{Morelli2024,Gamel2016,Wyderka2020,Shravan2024} (\textit{sector lengths}), pertaining to elements of the correlation tensor which describe a given $N$-particle correlations.
Even though sector lengths substantially compress the full information of the system’s state, they have found numerous applications, e.g., with regard to entanglement detection \cite{PhysRevA.91.042339,PhysRevA.101.012341}, monogamy relations \cite{PhysRevLett.114.140402}, and quantum coding theory \cite{gottesman1997}.
In our work, we also employ Bloch vector lengths, which, on the one hand are directly measurable in experiments, and on the other, due to their invariance under local unitary transformations, are particularly relevant to the field of randomized measurements, which has recently gained considerable attention \cite{CIESLINSKI20241,Elben2023,doi:10.1126/science.aau4963,PhysRevA.92.050301,Knips2020momentrandom,PhysRevLett.122.120505}.

Our analysis focuses on the evolution of a system governed solely by two-body interactions, leaving for future work the question of whether the emergence of genuinely three-body interactions could lead to qualitatively different outcomes. In particular, the types of interactions considered include general anisotropic spin exchange couplings, antisymmetric exchange Dzyaloshinskii-Moriya (DM) interactions, as well as Kaplan-Shekhtman-Entin-Wohlman-Aharony (KSEA) interactions.
The influence of different types of interactions and its interplay with the external fields were studied primarily in the context of geometry of states and the change of entanglement properties and also the influence on two-qubit gates in spin based quantum computer architectures (see e.g. \cite{Wang2005,Kuzmak2016,Frydryszak2019,Fedorova2021,ZHANG2007136,Houca2022}). Crucially, it is known that any arbitrary two-qubit transformation can be generated by a sequential combination of single-qubit rotations with an appropriate two-qubit interaction \cite{Zhang2003}, which in turn can be effectively simulated experimentally with the use of the same physical hardware \cite{Kapit2015}. Also recently, a class of two-body interactions was studied with the focus on determining the fastest generation of multipartite entanglement of different types \cite{PC2023}.

The paper is organized as follows. In Sec. \ref{sec2} we introduce the formalism and derive equations of motion for correlation tensor elements. We examine specific models of interaction of two qubits, deriving characteristic frequencies which determine the time evolution of correlations. In Sec. \ref{sec3} we focus on the stationarity of correlations under specific types of interactions. To this end, we perform a full characterization of the stationary density matrices corresponding to particular model Hamiltonians. In Sec. \ref{sec4} we analyse the evolution of correlations of three qubits under the influence of 2-body interactions. We visualize the trajectories of the correlation vectors in the corresponding correlation space and compare the observed features with the two-qubit case.

\section{Two-qubit correlation tensor evolution}
\label{sec2}

We begin our considerations with the system composed of two 2-level quantum subsystems, the state of which can be given by the density matrix, which can be written in the Bloch representation as
\ben\label{stan}
\rho = \frac14 \sum_{\mu,\nu}T_{\mu\nu}\sigma_{\mu}\otimes\sigma_{\nu},
\een
where the Greek indices $\mu,\nu\in\{0,1,2,3\}$,  $\sigma_{\mu}$ form an extended set of Pauli matrices, with $\sigma_0 \equiv \mathds{1}$ being a $2\times2$ identity matrix. The coefficients $T_{\mu\nu}$ are the elements of the \textit{correlation tensor}, defined as
\ben
T_{\mu\nu} = \Tr(\sigma_{\mu}\otimes\sigma_{\nu})\rho,
\een
the evolution of which we intend to analyze.

First, let us introduce the notation which will be used throughout the paper. The product of two Pauli matrices is given by
\ben
\sigma_i \sigma_j =\delta_{ij}\sigma_0 + i \varepsilon_{ijk}\sigma_k,
\een
where the Latin indices $i,j,k\in\{1,2,3\}$, $\delta_{ij}$ is the Kronecker delta, $\varepsilon_{ijk}$ is the Levi-Civita symbol, and additionally the summation over repeated indices is assumed. To make the subsequent notation concise, we need to introduce their Greek-indiced counterparts, so that for extended set of Pauli matrices the following relations hold
\ben
\sigma_{\mu} \sigma_{\zeta} &=&(\theta_{\mu\zeta\alpha} + i \varepsilon_{\mu\zeta\alpha})\sigma_{\alpha},\\
\comm{\sigma_{\mu}}{\sigma_{\zeta}} &=& 2 i \varepsilon_{\mu\zeta\alpha} \sigma_{\alpha},
\een
where 
\ben
 \theta_{\alpha\beta\gamma} \equiv
\begin{cases} 
1 & \text{if one index is 0 and the other two are equal,} \\
0 & \text{otherwise,}
\end{cases}
\een
whereas $\varepsilon_{\alpha\beta\gamma}$ takes the same values as $\varepsilon_{ijk}$, or 0 whenever any Greek index is 0, as proposed in \cite{Gamel2016}.

To make the notation more concise we will denote two-qubit Pauli matrices as
\ben
\Sigma_{\mu  \nu} = \sigma_{\mu}\otimes\sigma_{\nu},
\een
for which their products satisfy
\ben
\Sigma_{\mu\nu}\Sigma_{\zeta\eta} &=& (\theta_{\mu\zeta\alpha} + i \varepsilon_{\mu\zeta\alpha})(\theta_{\nu\eta\beta} + i \varepsilon_{\nu\eta\beta}) \Sigma_{\alpha\beta},\\
\comm{\Sigma_{\mu\nu}}{\Sigma_{\zeta\eta}} &=& 2i (\theta_{\mu\zeta\alpha} \varepsilon_{\nu\eta\beta} + \varepsilon_{\mu\zeta\alpha} \theta_{\nu\eta\beta})\Sigma_{\alpha\beta}.
\een
With this in mind we can write the state of a two-qubit system \eqref{stan} as
\ben
\rho = \frac14 T_{\mu\nu}\Sigma_{\mu\nu}.
\een
Although the considerations carried out further, as well as the obtained results, are general for all 2-level systems, we will continue to use the terminology related to spin-$\frac12$ systems.

Let us consider a general Hamiltonian describing the interaction between two spin-$\frac12$ particles given by
\ben\label{genham}
H = - \frac{1}{2} \sum_{ij} J_{ij} \sigma_i \otimes \sigma_j 
- \frac{1}{2} (\vec{B}\cdot \vec{\sigma} \otimes \mathds{1} +   \mathds{1} \otimes \vec{B}\cdot \vec{\sigma}),
\een
where $J_{ij}$ parametrize the exchange interaction between $i,j$ (commonly denoted as $x,y,z$) spin components of the respective particles, while the second term describes the interaction of the spins with the external magnetic field $\vec{B}$.
In terms of multiqubit Pauli matrices, we can write
\ben\label{hamil2q}
H = -\frac12 J_{\zeta\eta} \Sigma_{\zeta\eta},
\een
where we set $J_{00} = 0$, while the components of the external magnetic field are given by $J_{i0}$ and $J_{0j}$.

The time dependence of the elements of correlation tensor $T$ can be inferred by evaluating the correlations for time-evolved system
\ben
T_{\mu\nu}(t) = \Tr \Sigma_{\mu\nu}\rho(t),
\een
or by solving the equations of motion for the observables $\Sigma_{\mu\nu}$, which read as
\ben
\dot{\Sigma}_{\mu\nu} = i[H,\Sigma_{\mu\nu}].
\een
For the present two-qubit system this gives a system of ODE in the form
\ben\label{generalODE}
\dot{\Sigma}_{\mu\nu} = M_{\mu\nu}^{\alpha\beta}\Sigma_{\alpha\beta},
\een
where the explicit form of the tensor $M_{\mu\nu}^{\alpha\beta}$ is given in the Appendix \ref{AP1}. The tensor form of the system can be reformulated in a vectorized form as
\ben\label{vecODE}
\dot{\vec{\Sigma}} =  M\vec{\Sigma},
\een
where
\ben\label{sigmaset}
\vec{\Sigma} &=& (\Sigma_i)_{i=4\mu+\nu} = (\Sigma_{00},\Sigma_{01},\Sigma_{02},\Sigma_{03},\Sigma_{10},...,\Sigma_{33}),\hspace{0.55cm} \\
M &=& (M_{ij})_{i=4\mu+\nu, j=4\alpha+\beta},
\een
and the system can be solved by diagonalizing $16\times16$ matrix $M$. For a two-qubit Hamiltonian \eqref{genham}, however, finding a general solution is impractical due to its complex structure, therefore we restrict ourselves to analyzing a few special cases. First, let us notice that upon the exchange of indices $(\mu,\nu)\leftrightarrow(\alpha,\beta)$ in Levi-Civita symbols the tensor $M_{\mu\nu}^{\alpha\beta}$ changes sign rendering the tensor skew-symmetric, wchich applies to the matrix $M$ as well. The eigenvalues of a real skew-symmetric matrix form a set of complex conjugate pairs $\pm i\omega_n$, hence the system will evolve according to the characteristic frequencies $\omega_n$. The solution of the system defines the time dependence of correlation tensor elements $T_{\mu\nu}(t)$ for a set of initial conditions given by
\ben
\Sigma_{\mu\nu}(0) = \langle \Sigma_{\mu\nu} \rangle_{\rho(0)} = \Tr \Sigma_{\mu\nu} \rho(0).
\een
At this point, we already have that $T_{00}(t)=1$ as the trace of density matrix $\rho$ does not change. Because we consider a system composed of 2-dimensional subsystems, the dimensionality of the vector $\vec{\Sigma}$ is even, hence the fact that $\dot{\Sigma}_{00}=0$ implies that at least one more eigenvalue of a real skew-symmetric matrix $M$ must be zero. This fact by itself, however, does not imply that some other correlation tensor elements must necessarily be constant.
Note also, that due to highly nontrivial dependence of the frequencies $\omega_n$ on Hamiltonian parameters, the time evolution of $T_{\mu\nu}(t)$, except for a few special cases, does not need to be strictly periodic, but can also exhibit quasiperiodic behaviour, as we will see later.

\subsection{Heisenberg XXX interaction model}

\begin{figure*}
    \centering
    \includegraphics[width=0.9\textwidth]{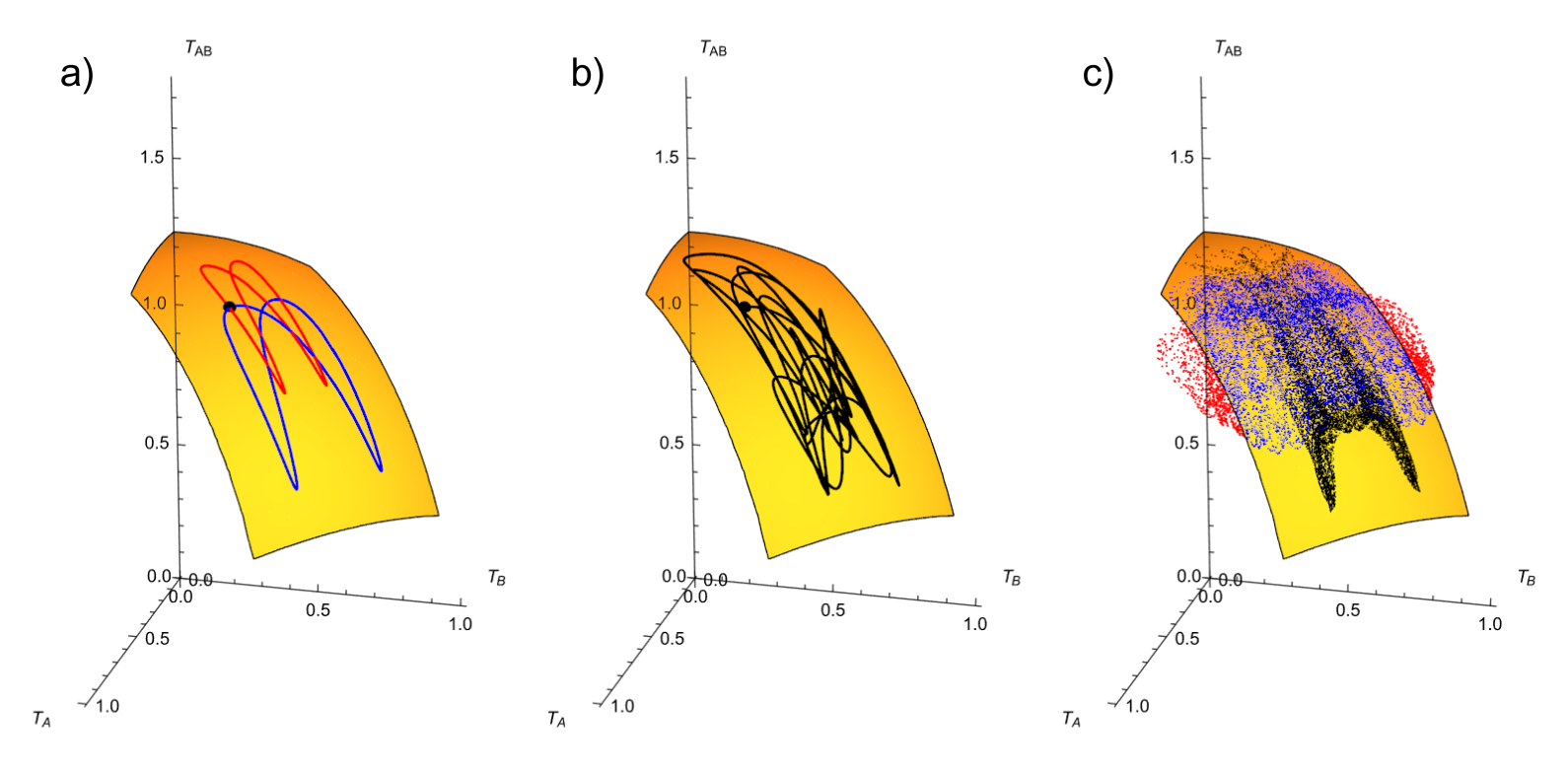}
    \caption{Trajectories of time evolved correlation vector $(T_A,T_B, T_{AB})$ generated by the XYZ Heisenberg interactions model. The evolution takes place on the isopuric surface pertaining to the purity of the initial state $\textrm{Tr}\rho_{\textrm{rand}}^2 = 0.6416$. a) Closed trajectories generated by the isotropic Heisenberg Hamiltonian with $J_1=J_2=J_3$, for the initial state $\rho_{\textrm{rand}}$ (blue) and $\tilde{\rho}_{\textrm{rand}}$ (red) which is obtained from $\rho_{\textrm{rand}}$ by local change of basis. Both states $\rho_{\textrm{rand}}$ and $\tilde{\rho}_{\textrm{rand}}$ share a common initial correlation vector (black dot), whereas the time evolution follow different paths; b) Trajectories generated by the anisotropic Heisenberg XYZ interaction with incommensurable parameters $J_1=\sqrt{3}, J_2=\sqrt{2}, J_3=\sqrt{5}$ (in arbitrary units). A similar irregular quasiperiodic type of trajectory is obtained in the presence of arbitrary external field $\vec{B}$; c) Comparison of time evolved (with discrete time steps) correlation vector for the initial state represented by density matrix $\rho_{\textrm{rand}}$ (black) and the initial unphysical state $\rho^{\Gamma_B}_{\textrm{rand}}$ (blue and red).}
    \label{flow}
\end{figure*}

Consider a model of interaction given by isotropic Heisenberg Hamiltonian with magnetic field $\vec{B}$:
\ben
J_{\zeta\eta} =
\begin{pmatrix}
0 & B_1 & B_2 & B_3 \\
B_1 & J & 0 & 0 \\
B_2 & 0 & J & 0 \\
B_3 & 0 & 0 & J
\end{pmatrix}.
\een
The characteristic frequencies are
\ben
\omega^{(2)} &=& |\vec{B}|, \\
\omega^{(1)} &=& 2|\vec{B}|, \\
\omega^{(1)} &=& 2J, \\
\omega^{(1)}_{\pm} &=& |\vec{B}|\pm 2J,\\
\omega^{(2)} &=& 0,
\een
where the superscript indicates the degree of degeneracy (formally, we also enumerate trivial frequencies, i.e., those equal to zero). For a field vector pointing at the direction of one axis, e.g. $\vec{B} = (0,0,B)$ we are able to explicitly write the general solution of the system, however, this does not provide much insight into the time behavior. In this aspect, we will illustrate the collective evolution of the correlation tensor elements using correlation model of two-qubit system, as presented in \cite{Morelli2024}. To this end, we evaluate components of the vector in two-qubit correlation space as
\ben
T_A = \sqrt{\sum_i T_{i0}^2}, \\
T_B = \sqrt{\sum_i T_{0i}^2}, \\
T_{AB} = \sqrt{\sum_{i,j} T_{ij}^2}. 
\een
Fig.\ref{flow} panel a) represents the time evolution of the vector $(T_A,T_B, T_{AB})$ for a random initial state, which we will denote as $\rho_{\textrm{rand}}$. We find that even though the general solution $T_{\mu\nu}(t)$ depends on the ratio $J/|\vec{B}|$, the correlation vector components $(T_A,T_B, T_{AB})$ do not depend on the field $\vec{B}$. At the same time the components of the correlation vector admit oscillatory behaviour, so that after a period $\frac{\pi}{2J}$ the vector reaches its initial position. Note that since the unitary evolution preserves the purity of the state, the length of the vector $(T_A,T_B, T_{AB})$ does not change, hence the evolution takes place on the surfaces of constant purity (which we will call isopuric surfaces) in the correlation model.

Bear in mind that the time evolution in the correlation model obviously does not represent Hamiltonian flow in phase space. Indeed, the mapping $\rho \mapsto (T_A,T_B, T_{AB})$ in not injective, and in principle infinitely many sets of $T_{\mu\nu}$ defining different states can be ascribed to a given vector $(T_A,T_B, T_{AB})$. This fact is illustrated in Fig.\ref{flow} panel a), where the same interaction parameters lead to significantly distinguishable trajectories corresponding to two different states with the same initial vector $(T_A,T_B, T_{AB})$. For a given randomly generated state ($\rho_{\textrm{rand}}$) with the initial set $T_{\mu\nu}$, we construct its correlation twin ($\tilde{\rho}_{\textrm{rand}}$) by the appropriate local cyclical change of (Latin) basis $T_{i,j} \rightarrow T_{i+1,j}$ which amounts to local unitary rotation around the axis $(1,1,1)$.

\subsection{Heisenberg XYZ interaction model}

Consider a model of interaction given by anisotropic Heisenberg Hamiltonian with magnetic field $\vec{B}$:
\ben\label{hamani}
J_{\zeta\eta} =
\begin{pmatrix}
0 & B_1 & B_2 & B_3 \\
B_1 & J_1 & 0 & 0 \\
B_2 & 0 & J_2 & 0 \\
B_3 & 0 & 0 & J_3
\end{pmatrix}.
\een
In the absence of magnetic field, $\vec{B}=0$, the characteristic frequencies become
\ben
\omega^{(1)}_{\pm} &=& J_1\pm J_2, \\
\omega^{(1)}_{\pm} &=& J_2\pm J_3, \\
\omega^{(1)}_{\pm} &=& J_1\pm J_3, \\
\omega^{(2)} &=& 0,
\een
and a general solution can be written explicitly. In contrast to the isotropic spin interaction, in the present case a qualitative difference comes into play. In general, the correlation tensor elements depend on three frequencies, and so do the components of the vector $(T_A,T_B, T_{AB})$. In the case when the frequencies form an incommensurable set the time dependence ceases its strictly periodic character, so that the correlation vectors no longer trace out closed curves on the isopuric surfaces, which is illustrated in Fig.\ref{flow} panel b). Even though the time evolution resembles a random walk on the isopuric surface, the correlation vector does not explore all accessible regions, even after a sufficiently long time. Note also, that as the purity of the initial state increases (which in turn decreases the area of isopuric surface), so does the accessible range for the trajectory of time evolved correlation vector.

In general, the time evolution of the vector $(T_A,T_B, T_{AB})$ is confined to the isopuric surface in the section of the correlation space defined by the region \cite{Morelli2024}
\ben\label{ogry}
T^2_{AB} \leq 3+ T^2_A + T^2_B - 4 T_A T_B - 4|T_A - T_B|.
\een
As said, a given triple $(T_A,T_B, T_{AB})$ fulfilling the above might not in general represent a correlation vector of a valid physical state. It becomes particularly evident if we consider the time evolution of $(T_A,T_B, T_{AB})$ pertaining to a nonphysical state given by, e.g., $\rho^{\Gamma_B}_{\textrm{rand}}$, where $\Gamma_B$ stands for partial transposition w.r.t. the subsystem $B$, which can be obtained by transformation $T_{\mu 2} \rightarrow -T_{\mu 2}$. In this case the time evolved triple $(T_A,T_B, T_{AB})$ no longer is confined to the correlation space given by \eqref{ogry} (see Fig.\ref{flow} panel c)).

In the presence of magnetic field parallel to the $z$-axis, 
$\vec{B}=(0,0,B)$, the characteristic frequencies are
\ben
\omega_1^{(1)} &=& J_1+J_2, \\
\omega_2^{(1)} &=& \sqrt{4B^2 + (J_1-J_2)^2}, \\
\omega^{(1)}_{\pm\pm} &=& |\frac{\omega_1 \pm \omega_2}{2} \pm J_3|, \\
\omega^{(2)} &=& 0.
\een
While the time evolution can exhibit similar quasiperiodic characteristics to the previous case ($\vec{B}=0$), the presence of magnetic field affects the states that are initially characterized by a high degree of local randomness. It can be analyzed by considering a class of \textit{Bell diagonal states}, the only nonzero correlation tensor elements of which are strictly diagonal $T_{\mu\mu}$. Analyzing the equations of motion \eqref{generalODE} for a general Heisenberg Hamiltonian with arbitrary vector of magnetic field we have the following:
\ben
\dot{T}_{mn} &=& J_m \varepsilon_{mni}T_{0i} +J_n \varepsilon_{nmi}T_{i0}\\
&& + B_j  \varepsilon_{jmi}T_{in} + B_j  \varepsilon_{jni}T_{ni}, \nonumber\\
\dot{T}_{m0} &=& J_i \varepsilon_{imj}T_{ji} + B_j  \varepsilon_{jmi}T_{i0}, \\
\dot{T}_{0n} &=& J_i \varepsilon_{inj}T_{ij} + B_j  \varepsilon_{jni}T_{0j}.
\een
We see that if the system being initially in Bell diagonal state undergoes the evolution, it will not change in terms of the diagonal elements $T_{\mu\mu}$ (and in fact all the others), which demonstrates that they form a class of stationary states of Heisenberg Hamiltonian in the absence of magnetic field (which is no longer the case for different type of interactions as we see in the next subsection).

The presence of the external field $\vec{B}$ affects the evolution of the correlation vector also with respect to the possibility of generating high degree of entanglement of the system considered. For this instace, consider an initial state of two qubits given by $|00\rangle$ where $|0\rangle$ is the eigenvector of $\sigma_3$ corresponding to the eigenvalue $+1$. The evolution of the pure state in the correlation space takes place exclusively on the curve
\ben
T^2_A+T^2_B+T^2_{AB} =3,
\een
with $T_A=T_B$, where the correlation vector $(0,0,\sqrt3)$ corresponds to a maximally entangled state.

In the absence of the magnetic field the system can be solved analytically, yielding
\ben
T^2_{AB}(t)=1 + 2 \sin^2(J_1-J_2)t,
\een
hence the system can evolve into a maximally entangled state. Consider now the presence of magnetic field $\vec{B}=(0,0,B)$ and the interaction parameters $J_1=-J_2=1$ in the units of $B$. Now we get
\begin{equation}
\max T^2_{AB} = 
\begin{cases}
3 & \text{for } 0 \leq B \leq 1, \\
1 + 2\left( \dfrac{2B}{1 + B^2} \right)^2 & \text{for } B > 1,
\end{cases}
\end{equation}
which, for $B>1$, corresponds to Schmidt coefficients $\{B/\sqrt{1+B^2},\;1/\sqrt{1+B^2}\}$,
so that a high degree of magnetic field prevents the system from evolving into a maximally entangled state.

\subsection{Antisymmetric exchange interactions models}

First, we consider a system with multicomponent Dzyaloshinskii–Moriya (DM) interaction model given by:
\ben\label{hamDM}
J_{\zeta\eta} =
\begin{pmatrix}
0 & B_1 & B_2 & B_3 \\
B_1 & 0 & D_3 & -D_2 \\
B_2 & -D_3 & 0 & D_1 \\
B_3 & D_2 & -D_1 & 0
\end{pmatrix}.
\een

\begin{figure*}[t]
    \centering
    \includegraphics[width=\textwidth]{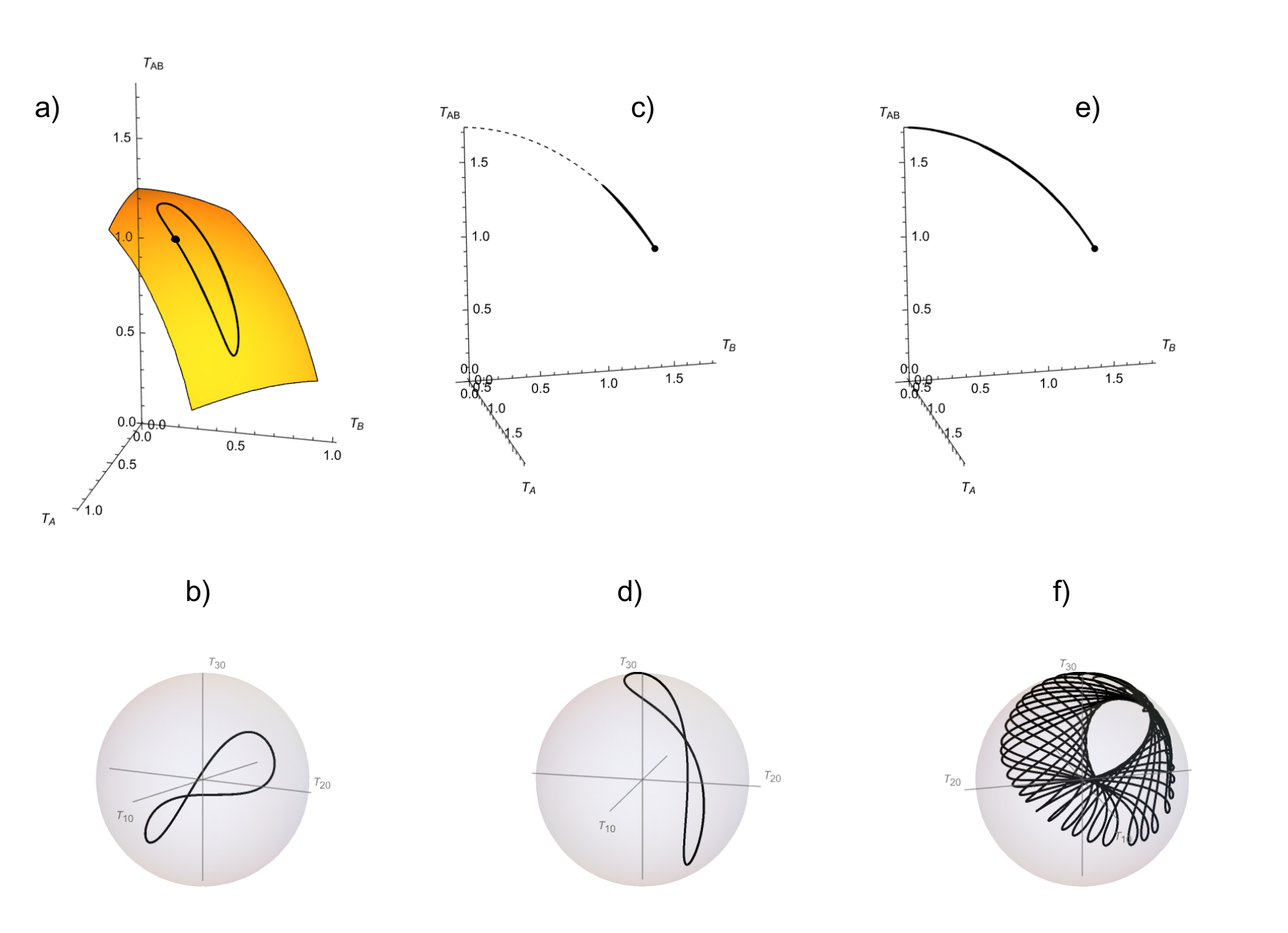}
    \caption{The evolution of the correlation vector generated by DM-type interactions with $D_1=D_2=D_3$ for different initial states in the correlation space (upper row) and the single qubit Bloch space (lower row). a) and b): Periodic evolution for random density matrix $\rho_{\textrm{rand}}$ in the absence of magnetic field. Note that the trajectory in the Bloch sphere does not intersect. A nonzero magnetic field $\vec{B}$ will make the corresponding trajectories non-periodic; c) and d): Periodic evolution for the initial state $|00\rangle$. $T^2_{AB}$ is upper-bounded by $\frac{17}{9}$. Note that the evolution in the Bloch sphere is uniplanar; e) and f): Quasiperiodic evolution for the initial state $|00\rangle$. A nonzero magnetic field $\vec{B}=(0,0,1)$ (in the units of $D$) renders $T^2_{AB}(t)$ to be a non-strictly-periodic function of $t$, the value of which can be arbitrarily close to 3.}
    \label{DM3x}
\end{figure*}

In the present case, the characteristic frequencies are
\ben
\omega^{(4)} &=& |\vec{D}|, \\
\omega^{(1)} &=& 2|\vec{D}|, \\
\omega^{(3)} &=& 0.
\een
Due to commensurability of the frequencies, the time behaviour of the correlation vector is strictly periodic, which for a random initial state is visualized in Fig.\ref{DM3x} (panels a) and b)).

For the initial state of the system given by $|00\rangle$ and the interaction parameters $D_1=D_2=D_3=D$ the system can be solved analytically, yielding
\ben
T^2_{AB}(t)=\frac19 (12 - 4 \cos 2 \sqrt3Dt + \cos 4 \sqrt3 Dt),
\een
the maximum value of which is
\ben
\max T^2_{AB} = \frac{17}{9},
\een
which corresponds to Schmidt coefficients $\sqrt{\frac12 \pm\frac{\sqrt5}{6}}$ (see Fig.\ref{DM3x}, c) and d)). Notably, however, for $D_2=D_3=0$ as well as for $D_1=D_3=0$ we get $\max T^2_{AB}=3$, whereas for $D_1=D_2=0$ the vector state of the system being initially at $|00\rangle$ does not evolve, hence $T^2_{AB}=1$.

The presence of the external field changes the picture considerably, where the characteristic frequencies of the system become
\ben
\omega^{(2)}_{\pm} &=& |\vec{D} \pm \vec{B}|, \\
\omega^{(1)}_{\pm} &=& \sqrt2 \sqrt{|\vec{D}|^2 + |\vec{B}|^2 \pm |\vec{D} - \vec{B}||\vec{D} + \vec{B}|}, \\
\omega^{(2)} &=& 0,
\een
and the trajectories of the correlation vectors no longer follow closed paths (with the exception of pure states for which the entire evolution in the correlation space is restricted to a single line). Notably, the presence of the external field enables the system being initially in the state $|00\rangle$ to evolve arbitrarily close to a maximally entangled state (see Fig.\ref{DM3x}, panels e) and f)).

Consider a system in an external magnetic field with multicomponent Kaplan-Shekhtman-Entin-Wohlman-Aharony (KSEA) interaction model given by:
\ben\label{hamKS}
J_{\zeta\eta} =
\begin{pmatrix}
0 & B_1 & B_2 & B_3 \\
B_1 & 0 & K_3 & K_2 \\
B_2 & K_3 & 0 & K_1 \\
B_3 & K_2 & K_1 & 0
\end{pmatrix}.
\een
In the present case, analytical diagonalization of the matrix $M$ is feasible in a simple case $K_1=K_2=K_3=K$ with vanishing external field, and yields
\ben
\omega^{(2)} &=& K, \\
\omega^{(1)} &=& 2K, \\
\omega^{(2)} &=& 3K, \\
\omega^{(3)} &=& 0.
\een

\begin{figure*}[t]
    \centering
    \includegraphics[width=\textwidth]{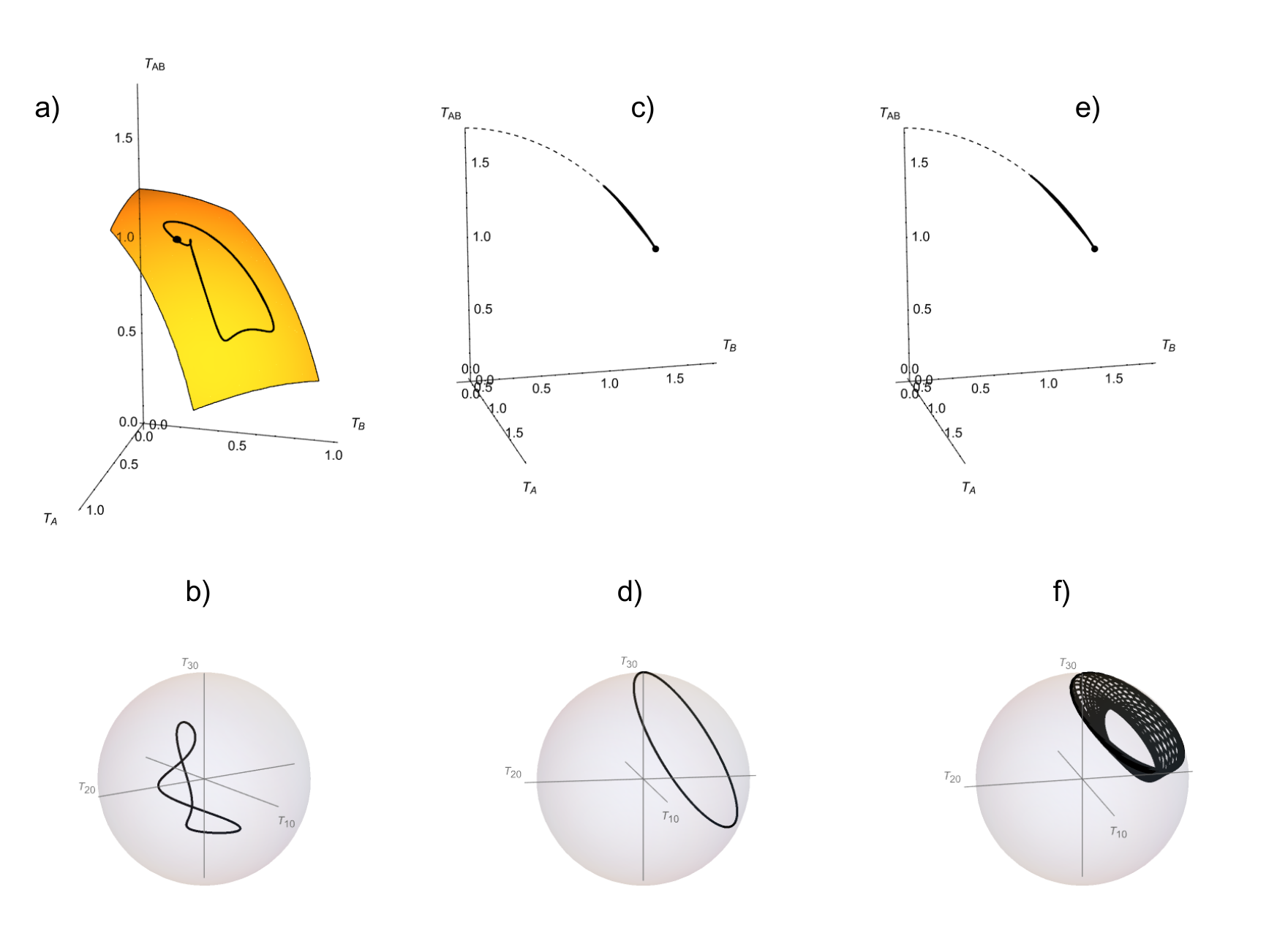}
    \caption{The evolution of the correlation vector generated by KSEA-type interactions with $K_1=K_2=K_3$ for different initial states in the correlation space (upper row) and the single qubit Bloch space (lower row). a) and b): Periodic evolution for random density matrix $\rho_{\textrm{rand}}$ in the absence of magnetic field. A nonzero magnetic field $\vec{B}$ makes the trajectories non-periodic; c) and d): Periodic evolution for the initial state $|00\rangle$. $T^2_{AB}$ is upper-bounded by $\frac{17}{9}$; e) and f): Quasiperiodic evolution for the initial state $|00\rangle$. A nonzero magnetic field $\vec{B}=(0,0,1)$ (in the units of $K$) renders $T^2_{AB}(t)$ to be a non-strictly-periodic function of $t$, the value of which is upper-bounded by $\approx 2.065$.}
    \label{KS3x}
\end{figure*}

A comparison of the KSEA-type with DM-type interactions yields the following.
For the initial state of the system given by $|00\rangle$ and the interaction parameters $K_1=K_2=K_3=K$ the system can be solved analytically, giving
\ben
T^2_{AB}(t)=\frac19 (13 - 4 \cos 6 Kt ),
\een
the maximum value of which, similarly to the previous case, is
\ben
\max T^2_{AB} = \frac{17}{9},
\een
which again corresponds to Schmidt coefficients $\sqrt{\frac12 \pm\frac{\sqrt5}{6}}$ (see Fig.\ref{KS3x}). Contrary to DM-type interactions, however, for all three unidirectional cases $K_1=K_2=0$, $K_1=K_3=0$ and $K_2=K_3=0$ we get $\max T^2_{AB}=3$.

\section{Two-qubit stationary correlations}
\label{sec3}

In this section, we will investigate spaces of stationary correlations pertaining to a given Hamiltonian. To this end, we analyze the nullspace of the matrix $M$, i.e.
\ben\label{kerr}
\textrm{ker}(M) = \{ \vec{S} \; |\; M \vec{S} = 0\},
\een
and note that the dimensionality of $\textrm{ker}(M)$ is strictly related to the number of trivial frequencies of the considered model.
Consider an operator $\rho$ given by the linear combination of $\Sigma_i$
\ben\label{const}
\rho=S_i \Sigma_i,
\een
where $S_i$ are components of an arbitrary vector from the nullspace $\textrm{ker}(M)$. By virtue of \eqref{vecODE} we can write
\ben
\dot{\rho} = S_i \dot{\Sigma}_i = S_i M_{ij} \Sigma_j,
\een
and due to \eqref{kerr} we have  $\dot{\rho}=0$, which means that the operator of the form \eqref{const} is a constant of motion. In order for the operator of this form to represent a valid density matrix, we must ensure its positive semidefiniteness, which can be achieved by demanding that the operator $\rho$ fulfills \cite{Gamel2016}
\ben
\textrm{Tr} \rho^2 &\leq& 1, \label{gamel1}\\
3\textrm{Tr} \rho^2  -2\textrm{Tr} \rho^3 &\leq& 1, \label{gamel2}\\
6\textrm{Tr} \rho^2  -8\textrm{Tr} \rho^3 + 6\textrm{Tr} \rho^4 - 3 (\textrm{Tr} \rho^2)^2 &\leq& 1. \label{gamel3}
\een
Although the nullspaces for cases of considered model Hamiltonians can be found, it is rather not instructive to present the general formulas due to their complex dependence on Hamiltonian parameters. Also, the density matrix parameters spaces, which define stationary correlations, can be neatly visualized for low dimensionality of the nullspaces, hence we will restrict the following analyses to a few representative examples. Here we also note that in the case of the most general form of Hamiltonian we always have 2 trivial frequencies $\omega=0$, hence the dimensionality of $\textrm{ker}(M)$ is 4.

\subsection{Heisenberg XYZ interaction model}

\begin{figure*}[t]
    \centering
    \includegraphics[width=0.85\textwidth]{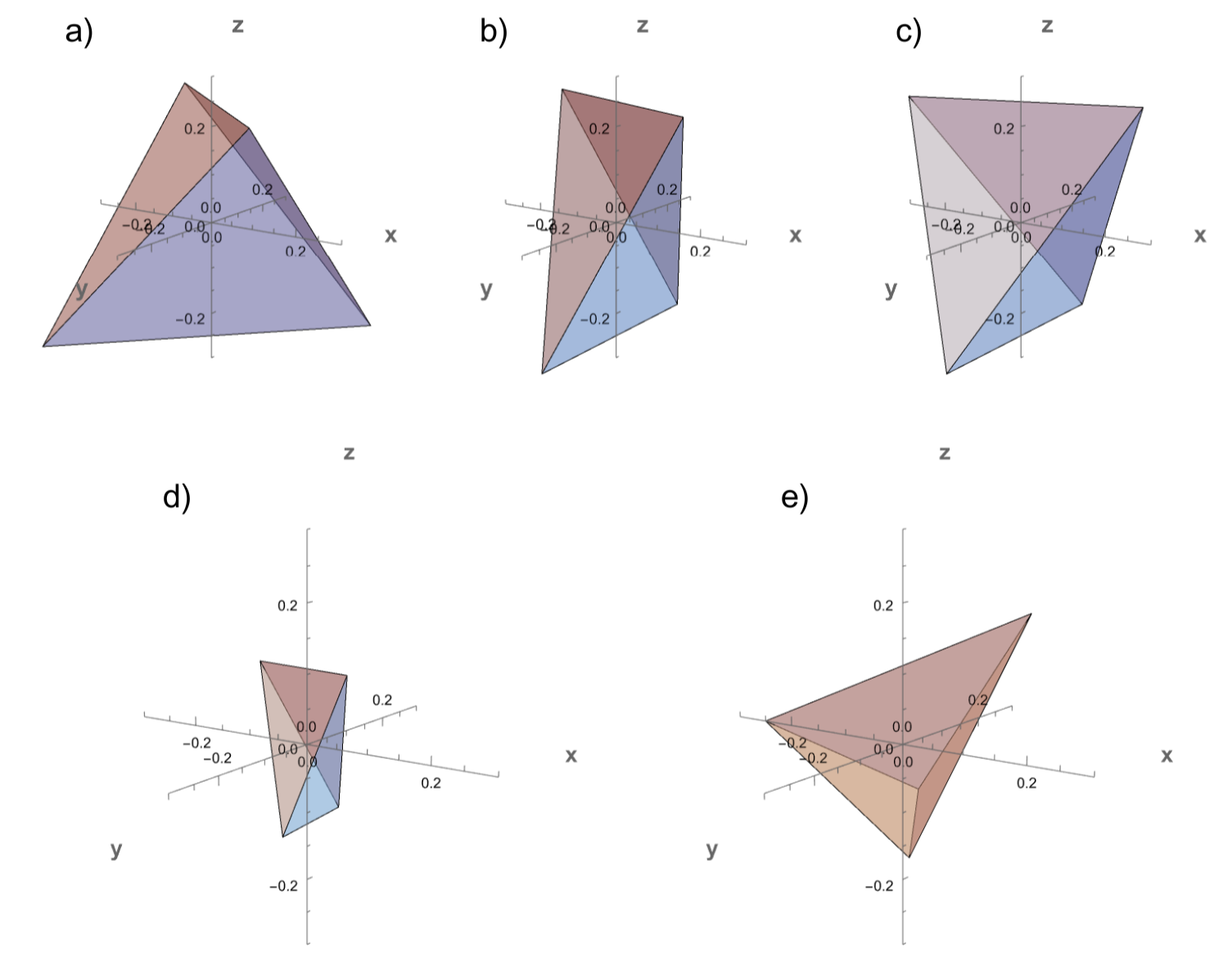}
    \caption{Tetrahedra representing the range of parameters defining stationary states for different model Hamiltonians: a) stationary correlations \eqref{rostatani0} for anisotropic Heisenberg Hamiltonian \eqref{hamani} with zero magnetic field; b) stationary correlations \eqref{rostatani} for anisotropic Heisenberg Hamiltonian \eqref{hamani} with the magnetic field for $\Delta=1$; c) stationary correlations \eqref{rostatani} for anisotropic Heisenberg Hamiltonian \eqref{hamani} with the magnetic field for $\Delta=-0.5$; d) stationary correlations \eqref{rostatDM} for Dzyaloshinskii-Moriya interaction model Hamiltonian \eqref{hamDM} with an external magnetic field; e) stationary correlations \eqref{rostatKS} for a model of KSEA interaction given by Hamiltonian \eqref{hamKS} with an external magnetic field.}
    \label{tetrahedrony}
\end{figure*}

Consider a model of interaction given by anisotropic Heisenberg Hamiltonian \eqref{hamani} with zero magnetic field $\vec{B}$. The nullspace of the operator $M$ is spanned by 4 vectors of the form
\ben
\vec{v}_0 &=& (1, 0, 0, 0, 0, 0, 0, 0, 0, 0, 0, 0, 0, 0, 0, 0), \nonumber\\
\vec{v}_1 &=& (0, 0, 0, 0, 0,
  1, 0, 0, 0, 0, 0, 0, 0, 0, 0, 0),  \nonumber\\
\vec{v}_2 &=& (0, 0, 0, 0, 0, 0, 0, 0, 0, 
  0, 1, 0, 0, 0, 0, 0),  \nonumber\\
\vec{v}_3 &=& (0, 0, 0, 0, 0, 0, 0, 0, 0, 0, 0, 0, 0, 0, 0, 1). \nonumber
\een
The vector $\vec{v}_0$ naturally corresponds to the operator $\Sigma_0\equiv\Sigma_{00}$, which renders the space of stationary correlations \eqref{const}, defined by the vector $\vec{S}=\frac14\vec{v}_0+x\vec{v}_1+y\vec{v}_2+z\vec{v}_3$, to be a 3-parameter family given by
\ben\label{rostatani0}
\rho &=& \frac14 \Sigma_{00} + x\Sigma_{11}+ y\Sigma_{22}+z\Sigma_{33}.
\een

Due to \eqref{gamel1}--\eqref{gamel3}, the set of parameters $x$, $y$ and $z$ defining the stationary density matrix is constrained by the respective set of three inequalities
\ben
\frac14 +4(x^2+y^2+z^2) &\leq& 1, \\
\frac58 +6(x^2+y^2+z^2) + 48xyz &\leq& 1, \\
\frac{29}{32}+3(x^2+y^2+z^2) - 24(x^4+y^4+z^4) &&\nonumber\\
+48(xyz+ x^2y^2+x^2z^2+y^2z^2) &\leq& 1.
\een

The first inequality constrains the set of parameters to a ball in the space $(x,y,z)$. The third inequality defines a region bounded by the surface of degree 4, which can be shown to be four distinct planes. These four planes cut out a tetrahedron whose vertices lie on the surface of the aforementioned ball. Ultimately, by restricting to the parameter space contained within the ball, it can be shown that the region defined by the second inequality is a convex domain that encloses the aforementioned tetrahedron. In conclusion, we have that the stationary density matrix pertaining to considered Hamiltonian is fully described by the family \eqref{rostatani0} with parameters $(x,y,z)$ lying inside the tetrahedron (see Fig.\ref{tetrahedrony} panel a)) the vertices of which are
\ben
&(-\frac14,\frac14,\frac14),& \nonumber \\
&(\frac14,-\frac14,\frac14),& \nonumber \\
&(\frac14,\frac14,-\frac14),& \nonumber \\
&(-\frac14,-\frac14,-\frac14),& \nonumber
\een
which, as expected, gives a family of Bell diagonal states.

Next, we consider anisotropic Heisenberg Hamiltonian \eqref{hamani} with magnetic field $\vec{B} = (0,0,B)$. In this case, the nullspace of the operator $M$ is spanned by 4 vectors of the form
\ben
\vec{v}_0 &=& (1, 0, 0, 0, 0, 0, 0, 0, 0, 0, 0, 0, 0, 0, 0, 0),  \nonumber\\
\vec{v}_1 &=& (0, 0, 0, 1, 0,
  \Delta, 0, 0, 0, 0, 0, 0, 1, 0, 0, 0),  \nonumber\\
\vec{v}_2 &=& (0, 0, 0, 0, 0, 1, 0, 0, 0, 
  0, 1, 0, 0, 0, 0, 0),  \nonumber\\
\vec{v}_3 &=& (0, 0, 0, 0, 0, 0, 0, 0, 0, 0, 0, 0, 0, 0, 0, 1), \nonumber
\een
where $\Delta=(J_x-J_y)/B$. Hence, the family of stationary matrices is given by
\ben\label{rostatani}
\rho &=& \frac14 \Sigma_{00} + x(\Sigma_{03}+\Sigma_{30}) \\ \nonumber
&& +(x\Delta+y)\Sigma_{11}+y\Sigma_{22}+z\Sigma_{33}.
\een

The set of parameters $x$, $y$ and $z$ defining the stationary density matrix is constrained by the respective set of three inequalities (given in Appendix \ref{AP2}). In this case, the first inequality constrains the set of parameters to an ellipsoid in the space $(x,y,z)$. The structure of the constraints is similar to the previous case (with additional Hamiltonian parameter $\Delta$), so that the same reasoning enables us to find the stationary density matrix pertaining to considered Hamiltonian, which is given by the family \eqref{rostatani} with parameters $(x,y,z)$ lying inside the tetrahedron (see Fig. \ref{tetrahedrony} panel b) for $\Delta=1$ and panel c) for $\Delta=-0.5$) the vertices of which are
\ben
&(0,\frac14,-\frac14),& \nonumber \\
&(0,-\frac14,-\frac14),& \nonumber \\
&(\frac{1}{2\sqrt{4+\Delta^2}},-\frac{\Delta}{4\sqrt{4+\Delta^2}},\frac14),& \nonumber \\
&(-\frac{1}{2\sqrt{4+\Delta^2}},\frac{\Delta}{4\sqrt{4+\Delta^2}},\frac14).& \nonumber
\een

It is also possible to analyze the most general case of Heisenberg Hamiltornian with arbitrarily oriented external field $\vec{B}$, where the nullspace of $M$ is again spanned by 4 vectors, leading to a 3-parameter family of stationary density matrices.

\subsection{Dzyaloshinskii–Moriya interaction model}

Consider a model of DM interaction given by \eqref{hamDM} with an external magnetic field $\vec{B}=(0,0,B)$. The nullspace of the operator $M$ is spanned by 4 vectors of the form
\ben
\vec{v}_0 &=& (1, 0, 0, 0, 0, 0, 0, 0, 0, 0, 0, 0, 0, 0, 0, 0),  \nonumber\\
\vec{v}_1 &=& (0, 0, 2, 0, 2, 1, 1, 2, 0, 1, 1, 0, 0, 0, 2, 0),  \nonumber\\
\vec{v}_2 &=& (0, -2, 0, 0, 0, 1,
   1, 0, -2, 1, 1, 2, 0, 2, 0, 0),  \nonumber\\
\vec{v}_3 &=& (0, 1, 1, 1, 1, 0, 1, 0, 1, -1, 0, 
  0, 1, 0, 0, 0). \nonumber
\een
In the present case, the family of stationary matrices is given by
\ben\label{rostatDM}
\rho &=& \frac14 \Sigma_{00}
+  (z-2y)\Sigma_{01}
+  (2x+z)\Sigma_{02}
+  z\Sigma_{03} \nonumber\\
&&+  (2x+z)\Sigma_{10}
+  (x+y)\Sigma_{11}
+  (x+y+z)\Sigma_{12}
+  2x\Sigma_{13} \nonumber\\
&&+  (z-2y)\Sigma_{20}
+  (x+y-z)\Sigma_{21}
+  (x+y)\Sigma_{22}
+  2y\Sigma_{23} \nonumber\\
&&+  z\Sigma_{30}
+  2y\Sigma_{31}
+  2x\Sigma_{32}.
\een

The set of parameters $x$, $y$ and $z$ defining the stationary density matrix is constrained by the respective set of three inequalities (given in Appendix \ref{AP2}). The geometrical structure of the constraints remains similar to the previous cases, hence the stationary density matrix pertaining to considered Hamiltonian is given by the family \eqref{rostatDM} with parameters $(x,y,z)$ lying inside the tetrahedron (see Fig.\ref{tetrahedrony} panel d)) the vertices of which are
\ben
&(1 +\sqrt2 -\sqrt3, -1 +\sqrt2 +\sqrt3, -4)/16\sqrt6,& \nonumber \\
&(1 -\sqrt2 +\sqrt3, -1 -\sqrt2 -\sqrt3, -4)/16\sqrt6,& \nonumber \\
&(-1 +\sqrt2 +\sqrt3, 1 +\sqrt2 -\sqrt3, 4)/16\sqrt6,& \nonumber \\
&(-1 -\sqrt2 -\sqrt3, 1 -\sqrt2 +\sqrt3, 4)/16\sqrt6.& \nonumber 
\een

Now, let us assume a model of DM interaction in the absence of external field, $\vec{B}=0$. In this case, the nullspace of the operator $M$ is spanned by the vectors
\ben
\vec{v}_0 &=& (1, 0, 0, 0, 0, 0, 0, 0, 0, 0, 0, 0, 0, 0, 0, 0),  \nonumber\\
\vec{v}_1 &=& (0, 0, 0, 0, 0, 0, 1, 0, 0, 1, 0, 0, 0, 0, 0, 1),  \nonumber\\
\vec{v}_2 &=& (0, 0, 0, 0, 0,  1, -1, 2, 0, 1, 1, 0, 0, 0, 2, 0),  \nonumber\\
\vec{v}_3 &=& (0, 0, 0, 0, 0, 0, 0, 1, 0, 0, 1,  0, 0, 1, 0, 0),  \nonumber\\
\vec{v}_4 &=& (0, 1, 1, 1, 1, 0, 0, 0, 1, 0, 0, 0, 1, 0, 0, 
  0),  \nonumber\\
\vec{v}_5 &=& (0, 0, 0, 0, 0, 1, 1, -2, 0, -1, -1, 2, 0, 0, 0, 0). \nonumber
\een
One can easily construct a stationary density matrix, which in this case is parametrized by 5 parameters, making further analysis much less transparent. 
In the currently considered case of a zero field, where no spatial direction is distinguished, the space of stationary correlation includes the correlation term $\Sigma_{33}$ (as the last nonzero coordinate of $\vec{v}_1$), as opposed to the earlier case with $z$-directional field. Hence, an initial state with arbitrary nonzero value of $\Sigma_{33}$ correlations which no longer belongs to the family of \eqref{rostatDM} is susceptible to evolution under DM interaction in the presence of the external $z$-directional field.

\subsection{KSEA interaction model}

Consider a model of KSEA interaction given by \eqref{hamKS} with an external magnetic field $\vec{B}=(0,0,B)$. The nullspace of the operator $M$ is spanned by 4 vectors of the form
\ben
\vec{v}_0 &=& (1, 0, 0, 0, 0, 0, 0, 0, 0, 0, 0, 0, 0, 0, 0, 0),  \nonumber\\
\vec{v}_1 &=& (0, 0, 0, 0, 0, 1, 0, 0, 0, 0, 1, 0, 0, 0, 0, 1),  \nonumber\\
\vec{v}_2 &=& (0, -1, -1, 0, -1, 
  1, 1, 1, -1, 1, 1, 1, 0, 1, 1, 0),  \nonumber\\
\vec{v}_3 &=& (0, 1, 1, 1, 1, -1, 0, 0, 1, 
  0, -1, 0, 1, 0, 0, 0). \nonumber
\een
In the present case, the family of stationary matrices is given by
\ben\label{rostatKS}
\rho &=& \frac14 \Sigma_{00}
+  (z-y)\Sigma_{01}
+  (z-y)\Sigma_{02}
+  z\Sigma_{03} \nonumber\\
&&+  (z-y)\Sigma_{10}
+  (x+y-z)\Sigma_{11}
+  y\Sigma_{12}
+  y\Sigma_{13} \nonumber\\
&&+  (z-y)\Sigma_{20}
+  y\Sigma_{21}
+  (x+y-z)\Sigma_{22}
+  y\Sigma_{23} \nonumber\\
&&+  z\Sigma_{30}
+  y\Sigma_{31}
+  y\Sigma_{32}
+  x\Sigma_{33}.
\een

The set of parameters $x$, $y$ and $z$ defining the stationary density matrix is constrained by the respective set of three inequalities (given in Appendix \ref{AP2}). Again, the geometrical structure of the constraints is similar as in the previous cases, hence the stationary density matrix pertaining to considered Hamiltonian is given by the family \eqref{rostatKS} with parameters $(x,y,z)$ lying inside the tetrahedron (see Fig.\ref{tetrahedrony} panel e)) the vertices of which are
\ben
&(-0.0232,0.0737,-0.0862),& \nonumber \\
&(0.1588,0.0764,0.1896),& \nonumber \\
&(0.1145,-0.1501,-0.1034),& \nonumber \\
&(-\frac14,0,0).& \nonumber
\een

In the absence of the external field, the nullspace of the operator $M$ is spanned by the vectors
\ben
\vec{v}_0 &=& (1, 0, 0, 0, 0, 0, 0, 0, 0, 0, 0, 0, 0, 0, 0, 0),  \nonumber\\
\vec{v}_1 &=& (0, 0, 0, 0, 0, 0, 1, 0, 0, 1, 0, 0, 0, 0, 0, 1),  \nonumber\\
\vec{v}_2 &=& (0, 0, 0, 0, 0, 1, 0, 0, 0, 0, 0, 1, 0, 0, 1, 0),  \nonumber\\
\vec{v}_3 &=& (0, 0, 0, 0, 0, -1, 1, 1, 0, 1, 0, 0, 0, 1, 0, 0),  \nonumber\\
\vec{v}_4 &=& (0, 1, 1, 1, 1, 0, 0, 0, 1, 0, 0, 0, 1, 0, 0, 0),  \nonumber\\
\vec{v}_5 &=& (0, 0, 0, 0, 0, -1, 1, 0, 0, 1, -1, 0, 0, 0, 0, 0), \nonumber
\een
so that a stationary density matrix is parametrized by 5 parameters, again constrained by \eqref{gamel1}--\eqref{gamel3}.

\section{Multiqubit correlation tensor evolution}
\label{sec4}

For a multiqubit system, the multiqubit Pauli matrices are denoted compactly as
\ben
\Sigma_{\mu ... \nu} = \sigma_{\mu}\otimes ... \otimes\sigma_{\nu}.
\een
The state of a $N$-qubit system can be written as
\ben
\rho = \frac{1}{2^N} T_{\mu ... \nu}\Sigma_{\mu ... \nu},
\een
and a general Hamiltonian describing the interaction between two spin-$\frac12$ particles written in terms of multiqubit Pauli matrices is
\ben
H = -\frac12 J_{\zeta ...\eta} \Sigma_{\zeta ... \eta}.
\een
Here we consider only 2-body interaction terms, so that in the term $J_{\zeta ...\eta}$ only two indices are nonzero, while the components of the external field are given by $J_{\zeta ...\eta}$ with only one nonzero Latin index. The remaining elements are set to 0.

We restrict further analysis to the system composed of three 2-dimensional subsystems.
For three qubits, the products and commutation relations of Pauli matrices satisfy
\ben
\Sigma_{\mu\nu\lambda}\Sigma_{\zeta\eta\omega} &=& (\theta_{\mu\zeta\alpha} + i \varepsilon_{\mu\zeta\alpha})  \\
&&(\theta_{\nu\eta\beta} + i \varepsilon_{\nu\eta\beta}) (\theta_{\lambda\omega\gamma} + i \varepsilon_{\lambda\omega\gamma}) \Sigma_{\alpha\beta\gamma},\nonumber\\
\comm{\Sigma_{\mu\nu\lambda}}{\Sigma_{\zeta\eta\omega}} &=& 2i (\theta_{\mu\zeta\alpha} \varepsilon_{\nu\eta\beta} \theta_{\lambda\omega\gamma} + \varepsilon_{\mu\zeta\alpha} \theta_{\nu\eta\beta} \theta_{\lambda\omega\gamma}  \\
&&+ \theta_{\mu\zeta\alpha} \theta_{\nu\eta\beta} \varepsilon_{\lambda\omega\gamma} - \varepsilon_{\mu\zeta\alpha} \varepsilon_{\nu\eta\beta} \varepsilon_{\lambda\omega\gamma})\Sigma_{\alpha\beta\gamma},\nonumber
\een
and the evolution of correlation tensor elements $T_{\mu \nu \lambda}$ subjected to a given Hamiltonian is given by a linear system
\ben
\dot{T}_{\mu \nu \lambda} = M^{\alpha \beta \gamma}_{\mu \nu \lambda} T_{\alpha \beta \gamma}.
\een
The system can be solved by means of diagonalization of the matrix $M^{\alpha \beta \gamma}_{\mu \nu \lambda}$ (the explicit formula of which is given in the Appendix \ref{AP1}).

To illustrate the obtained results, it is useful to introduce the length of correlation sectors defined as
\ben
A_1&=&\sum_{i}T^2_{i00} + T^2_{0i0} + T^2_{00i},\\
A_2&=&\sum_{i,j}T^2_{ij0} + T^2_{i0j} + T^2_{0ij},\\
A_3&=&\sum_{i,j,k}T^2_{ijk}.
\een
Naturally, for, e.g., a 2-qubit system we have $A_1=T^2_A+T^2_B$ and $A_2=T^2_{AB}$.
With this in mind, the evolution of the system can be tracked by a trajectory of a vector $(A_1,A_2,A_3)$ in the corresponding space \cite{Wyderka2020}, while the evolution takes place on an isopuric surface in the form of a plane described by the equation
\ben
A_1+A_2+A_3=8\textrm{Tr}\rho^2(0) - 1.
\een
A notable characteristic of this picture is the fact that for pure states we have $A_2=3$, hence the evolution takes place exclusively on the line $A_1+A_3=4$.

\begin{figure*}
    \centering
    \includegraphics[width=1\textwidth]{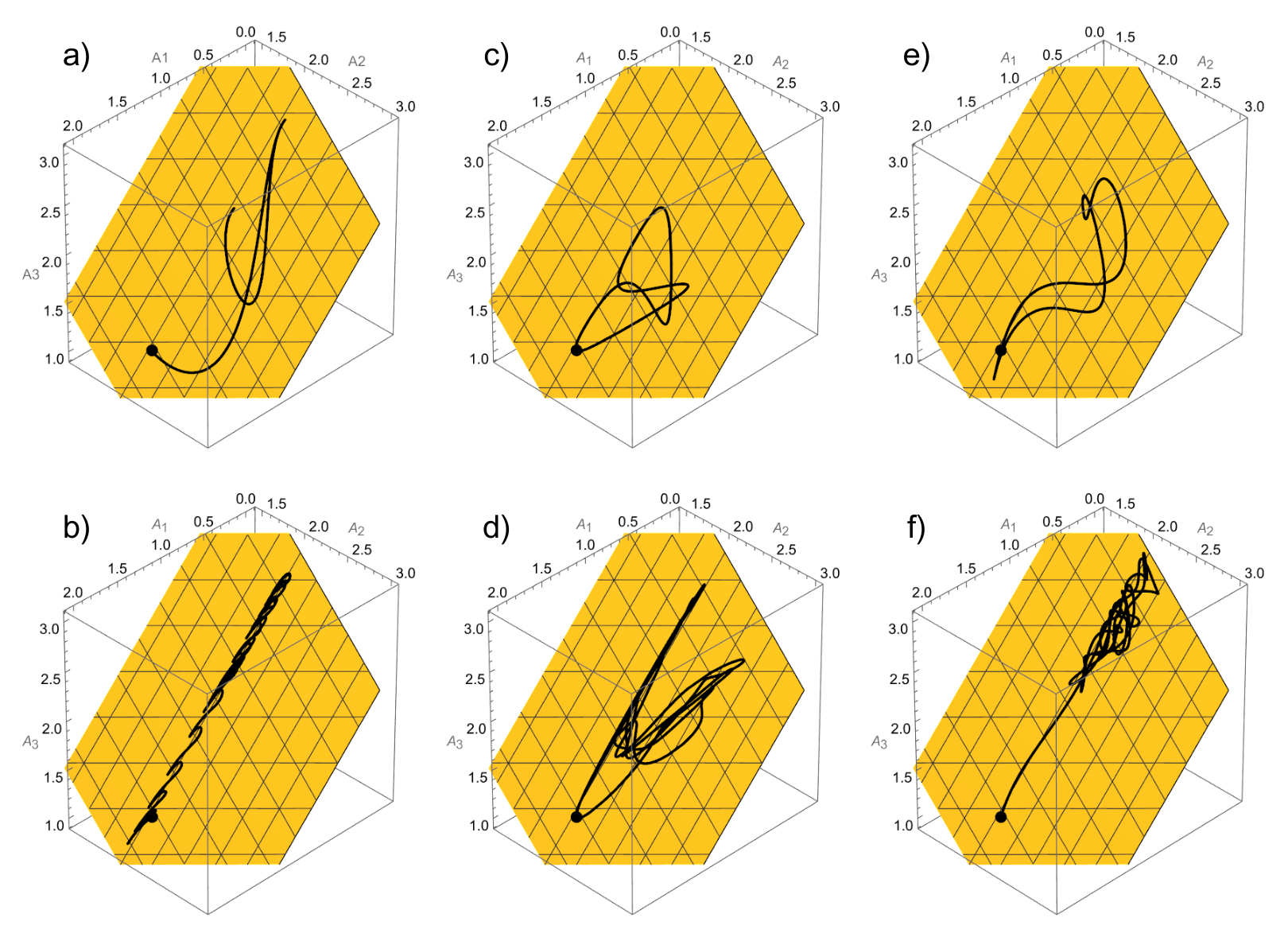}
    \caption{The evolution of the correlation vector $(A_1,A_2,A_3)$ generated by different types of interactions for a random density matrix. The initial state $\rho_{\textrm{rand3}}$ is chosen as a proper 3:1 mixture of two randomly generated pure states, with purity $\textrm{Tr}\rho^2_{\textrm{rand3}}=0.7575$. a) A periodic trajectory in a form of a line for Heisenberg type interaction with $\vec{J}=(1,-1,0)$ in the presence of the external field $\vec{B}=(0,0,1)$; b) A squeezed trajectory for $\vec{J}=(1,-1,0)$ with a strong external directional field $\vec{B}=(0,0,5)$; c) A closed periodic trajectory for DM-type interactions $\vec{D}=(1,1,1)$ without the external field and d) with a strong external directional field $\vec{B}=(0,0,5)$; e) A closed periodic trajectory for KSEA-type interactions $\vec{K}=(1,1,1)$ without the external field; f) A squeezed trajectory for $\vec{K}=(1,1,1)$ with a strong external directional field $\vec{B}=(0,0,5)$.}
    \label{3qu}
\end{figure*}

\begin{figure}
    \centering
    \includegraphics[width=1\linewidth]{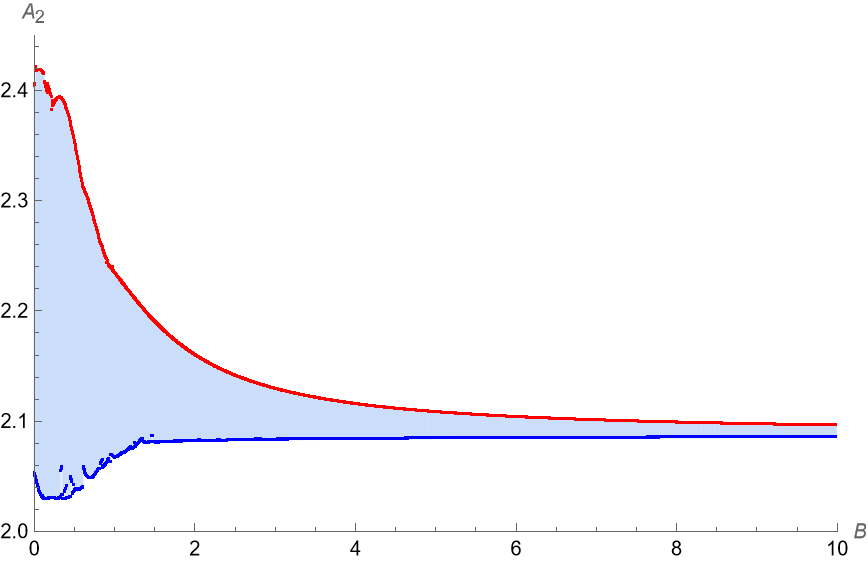}
    \caption{The range of the allowed value of 2-body correlation sector $A_2$ of a system described by the initial state $\rho_{\textrm{rand3}}$ interacting via DM+KSEA interactions (with arbitrary parameters, except $D_3=K_3=0$) with respect to the strength of the external directional field $\vec{B}=(0,0,B)$. For the initial state we have $A_2=2.0871$.}
    \label{maxmin}
\end{figure}

We observe that the 3- and 2-qubit systems share some common characteristics, which can be illustrated by trajectories in the correlation space obtained for a randomly selected initial state, as presented in Fig.\ref{3qu}. In the case when the system's characteristic frequencies form a commensurate set, the trajectory of the correlation vector becomes periodic, either in the form of a line or a closed loop. Yet, the system can exhibit quasiperiodic behavior even in the absence of the external field, as witnessed in the case of 2-qubit systems.

The mutual 2-body interactions between subsystems in general tend to increase the length of 3-partite correlation sector $A_3$. Notably, similarly as in the case of 2-qubit systems, emergence of strong external field for several classes of interaction types decreases the range of total 2-body correlations $A_2$ that the system could exhibit if subjected to mutual interactions without the external field. This becomes particularly evident in Heisenberg type interaction for a specific case $J_1=-J_2$, $J_3=0$ in the presence of a directional field $\vec{B}=(0,0,B)$ (see Fig.\ref{3qu} panels a) and b).
With this respect, antisymmetric type of interactions (DM as well as KSEA) provides more general observation. Consider three spins interacting solely via DM+KSEA interactions. Numerical analysis shows that whenever $i$-th component of considered types of interaction is nonexistent, the strong directional field in $i$-th direction forbids the change of total 2-body correlations given by the initial value $A_2$ (see Fig.\ref{maxmin}), even though no such restriction is witnessed for 3-body correlations $A_3$ (if the system is allowed to evolve sufficiently long). In the presence of strong external field, the evolution of the system effectively mimics the behaviour of pure states in the sense that ratio $A_3:A_1$ varies in time, while $A_2$ remains constant.

\section{Conclusions}

We analyzed the behavior of correlation tensor elements subjected to evolution generated by different classes of interactions. We focused solely on analyzing the effects of two-body interactions, taking into account an external field common to all interacting subsystems. Each set of interaction parameters defines a set of characteristic frequencies of the system, according to which the system evolves. The evolution can be compactly visualized by the trajectories of correlation vectors in the corresponding correlation space. Due to mutual commensurability of the aforementioned frequencies, the system exhibits either periodic or quasiperiodic behaviour.

For each model Hamiltonian, trivial (i.e. equal to zero) frequencies determine the space of stationary correlations, which describe a family of states whose density matrices commute with the given Hamiltonian. For the most general form of Hamiltonian, which includes all types of considered interaction, including the external field, the family of stationary states for two qubits is in general described by 3 free parameters, or 7 in the case of the system of three qubits. At this point, we stress that all the conclusions are drawn on the assumption that the Hamiltonians do not include 3-body interaction terms. This issue by itself requires further analysis, which was not undertaken in the present work.

We showed, that as the system subjected to interaction evolves, in certain cases a strong external field can significantly constrain the variability of system parameters. Although this observation pertains to the length of sector $A_2$, it remains an open question whether specific constraints may limit the variability of only certain elements of the correlation tensor $T_{\mu\nu}$ or $T_{\mu\nu\lambda}$. With this respect, we note the following: it can be shown that at least in certain cases the structure of the matrix $M$ causes the elements of the tensor $T$ to be grouped into sets of mutual dependencies. In the general case, however, a decomposition into mutually orthogonal subspaces is not possible, which implies that the evolution of a given tensor element depends on all the others, i.e., it necessitates knowledge of the complete state.
Lastly, the presented analysis naturally ascribes to the correlation tensor elements which uniquely define a given density matrix of an arbitrary mixed state (as, e.g., considered $\rho_{\textrm{rand3}}$ being a proper mixture of two random pure states). This raises the question of whether a similarly general pattern of variability will be observed in the case of improper mixtures. It is natural to expect that improper mixtures originating, for instance, from certain well-defined classes of multipartite states could display qualitatively distinct features compared to other types of states.

\section*{Acknowledgements}

We thank Paweł Cie\'sli\'nski and Krzysztof Szczygielski for fruitful comments and discussions. A.K. acknowledges the financial support from Gda\'nsk-Lund Cluster: Quantum Development Potential, funded by NAWA Strategic Partnership (BNI/PST/2023/1/00013/U/00001).

\onecolumngrid

\appendix

\section{Equations of motion for correlation tensor elements}\label{AP1}

For 2 qubits, the evolution of correlation tensor elements $T_{\mu \nu}$ subjected to the Hamiltonian \eqref{hamil2q} is given by
\ben
\dot{T}_{\mu \nu} = M^{\alpha \beta}_{\mu \nu} T_{\alpha \beta},
\een
where
\ben
M^{\alpha \beta}_{\mu \nu} = J_{\zeta \eta} (\theta_{\zeta \mu \alpha} \epsilon_{\eta \nu \beta} + \epsilon_{\zeta \mu \alpha} \theta_{\eta \nu \beta}).
\een
The interaction terms in considered model Hamiltonians are
\ben
J_{ij} = \delta_{ij}J_j
\een
for Heisenberg XYZ interaction model,
\ben
J_{ij} = \varepsilon_{ijk}J_{k}
\een
for DM interaction model, and
\ben
J_{ij} = \varepsilon^2_{ijk}J_{k}
\een
for KSEA interaction model.

For 3 qubits, the evolution of correlation tensor elements $T_{\mu \nu \lambda}$ subjected to a given Hamiltonian is given by
\ben
\dot{T}_{\mu \nu \lambda} = M^{\alpha \beta \gamma}_{\mu \nu \lambda} T_{\alpha \beta \gamma},
\een
where
\ben
M^{\alpha \beta \gamma}_{\mu \nu \lambda} = J_{\zeta \eta \omega} (\epsilon_{\zeta \mu \alpha} \theta_{\eta \nu \beta} \theta_{\omega \lambda \gamma} + \theta_{\zeta \mu \alpha} \epsilon_{\eta \nu \beta} \theta_{\omega \lambda \gamma} +  \theta_{\zeta \mu \alpha} \theta_{\eta \nu \beta} \epsilon_{\omega \lambda \gamma} - \epsilon_{\zeta \mu \alpha} \epsilon_{\eta \nu \beta} \epsilon_{\omega \lambda \gamma}).
\een

\section{Stationary correlations}\label{AP2}

Below, we write the explicit form of inequalities \eqref{gamel1}-\eqref{gamel3} which define the region of parameters that form the family of stationary states subjected to a given Hamiltonian.

The family of stationary states
\ben
\rho &=& \frac14 \Sigma_{00} + x(\Sigma_{03}+\Sigma_{30})  +(x\Delta+y)\Sigma_{11}+y\Sigma_{22}+z\Sigma_{33}
\een
for anisotropic Heisenberg Hamiltonian \eqref{hamani} with magnetic field $\vec{B} = (0,0,B)$, is fully described by the set of parameters that are confined by
\ben
\frac{1}{4} + 4 \big( y^2 + z^2 \big) + 4 x^2 \big( 2 + ( y + \Delta) ^2 \big) &\leq& 1, \\
\frac{5}{8} + 6 y^2 + 6\Big( z^2 + 8xyz \big(y + \Delta \big)+ 
x^2 \big( 2 - 8z + ( y + \Delta ) ^2 \big)\Big) &\leq& 1,\\
\frac{29}{32} + 3y^2 + 3\Big( z^2 - 8\big(y^2 - z^2\big)^2 - 64x^3y\big(y + \Delta \big) + 16xyz \big(y + \Delta \big) - 8x^4 \big(y + \Delta \big)^2 \big(4 + (y + \Delta)^2\big) && \nonumber\\
+ x^2 \big(2 + 16y^4 + 32y^3\Delta + \Delta^2 + 2y(\Delta + 16z^2\Delta)  + y^2(-31 + 16z^2 + 16\Delta^2)
\big) 
+ 16z\big(-1 + z(2 + \Delta^2)\big) \Big) &\leq& 1,
\een
where $\Delta=(J_x-J_y)/B$. 
The first and third inequalities together define a tetrahedron presented in Fig.\ref{tetrahedrony}, panels b) and c).

The family of stationary states
\ben
\rho &=& \frac14 \Sigma_{00}
+  (z-2y)\Sigma_{01}
+  (2x+z)\Sigma_{02}
+  z\Sigma_{03}+  (2x+z)\Sigma_{10}
+  (x+y)\Sigma_{11}
+  (x+y+z)\Sigma_{12}
+  2x\Sigma_{13} \nonumber\\
&&+  (z-2y)\Sigma_{20}
+  (x+y-z)\Sigma_{21}
+  (x+y)\Sigma_{22}
+  2y\Sigma_{23} +  z\Sigma_{30}
+  2y\Sigma_{31}
+  2x\Sigma_{32}
\een
for a model of DM interaction given by \eqref{hamDM} with an external magnetic field $\vec{B}=(0,0,B)$, is fully described by the set of parameters that are confined by
\ben
\frac{1}{4} + 16 \big(5x^2 + 5y^2 -2 yz +2 z^2 + 2x (y+z)\big) &\leq& 1,\hspace{0.8cm} \\
\frac{5}{8} + 24 \big(x^2 (5-48z) + 2z^2 - 2yz (1+12z) + y^2 (5+48z) + 2x (y+z-12 z^2)\big) &\leq& 1, \\
\frac{29}{32} + 12\Big(5x^2 - 32 x^4 + 2xy - 640 x^3 y + 5y^2 - 3264 x^2 y^2 -640 x y^3 - 32 y^4 + 2 \big(x-y\big) \big(1+64 x^2  && \nonumber\\
+ 16 y (-3 + 4y)+ 16 x (-3 + 40 y)\big) z + 2\big(1-24 y + 24(x(-1 +8x) + 16 xy + 8 y^2)\big) z^2 + 128 \big(x-y\big) z^3 - 32 z^4\Big) &\leq& 1.
\een
The first and third inequalities together define a tetrahedron presented in Fig.\ref{tetrahedrony}, panel d).

The family of stationary states
\ben
\rho &=& \frac14 \Sigma_{00}
+  (z-y)\Sigma_{01}
+  (z-y)\Sigma_{02}
+  z\Sigma_{03} +  (z-y)\Sigma_{10}
+  (x+y-z)\Sigma_{11}
+  y\Sigma_{12} \nonumber\\
&& +  y\Sigma_{13}+  (z-y)\Sigma_{20}
+  y\Sigma_{21}
+  (x+y-z)\Sigma_{22}
+  y\Sigma_{23} +  z\Sigma_{30}
+  y\Sigma_{31}
+  y\Sigma_{32}
+  x\Sigma_{33}
\een
for a model of KSEA interaction given by \eqref{hamKS} with an external magnetic field $\vec{B}=(0,0,B)$, is fully described by the set of parameters that are confined by
\ben
\frac{1}{4} + 4\big(3x^2 + 4xy + 12y^2 - 4(x + 3y) z + 8z^2\big) &\leq& 1, \\
\frac{5}{8} + 6 \Big(8x^3 + x^2 (3 + 16y - 16z)- 4x \big(y (-1 + 8y) + z - 4yz + 4 z^2\big) && \nonumber\\
+ 4 \big(-8 y^3 + 2 z^2 (1+2z) - 3yz (1 + 8z)+ y^2 (3 + 32z)\big)\Big) &\leq& 1, \\
\frac{29}{32} + 3 \Big(24 x^4 - 128 y^4 + 16 x^3 \big(1 + 4y - 4z\big) - 64 y^3 \big(1 + 12 z\big) + 8 z^2 \big(1 + 4 (1 - 4z) z\big) &&\nonumber\\
- 4x \big(y (-1 + 16 y(1+8y)) + z + 8y (-1 + 24 y) z + 8 (1 - 16 y) z^2 - 96 z^3 \big) + 4 y^2 \big(3 + 64 z (1 + 6z)\big)  &&\nonumber\\
-4 yz \big(3 + 16 z (3 + 8z)\big) + x^2 \big(3 - 488 y^2 - 32z (1 + 8z) + 32 y (1 + 10 z)\big)\Big) &\leq& 1.
\een
The first and third inequalities together define a tetrahedron presented in Fig.\ref{tetrahedrony}, panel e).
The coordinates of the tetrahedron vertices, which are given with approximations in the main text, are the solutions of qubic equations
\ben
2368x^3-592x^2+28x+1&=&0, \\
1184y^3-20y+1&=&0, \\
592z^3-16z-1&=&0.
\een

\section{Random states used in simulations}

Below we present the forms of the initial states used in the simulations.

The 2-qubit state $\rho_{\textrm{rand}}$ has been obtained with the use of random complex $4\times4$ matrix $A$
\begin{equation}
    \rho_{\textrm{rand}}= A^{\dagger}A/\textrm{Tr}A^{\dagger}A =  \frac14 T_{\mu\nu}\Sigma_{\mu\nu},
\end{equation}
giving the set of correlation tensor elements which uniquely characterize the state:
\begin{equation}
\begin{aligned}
T_{00} &= 1,   & T_{01} &= 0.1465,   & T_{02} &= -0.2315,  & T_{03} &= 0.1494, \\
T_{10} &= -0.0219,  & T_{11} &= -0.2760,  & T_{12} &= 0.5326,   & T_{13} &= 0.1250, \\
T_{20} &= -0.4355,  & T_{21} &= -0.3554,  & T_{22} &= 0.1621,   & T_{23} &= -0.5886, \\
T_{30} &= -0.0804,  & T_{31} &= 0.5104,   & T_{32} &= 0.2769,   & T_{33} &= -0.2463.
\end{aligned}
\end{equation}

The 3-qubit state $\rho_{\textrm{rand3}}$ has been obtained as a mixture of 2 randomly generated pure states
\begin{equation}
    \rho_{\textrm{rand3}}= \frac34 |\Psi\rangle\langle\Psi|+ \frac14 |\Phi\rangle\langle\Phi|=  \frac18 T_{\mu\nu \lambda}\Sigma_{\mu\nu\lambda}.
\end{equation}
The chosen canonical form of pure 3-qubit states renders selected correlation tensor elements zero, while the set fully characterizes the state
\begin{equation}
\begin{aligned}
T_{000} &= 1,        & T_{001} &= 0.5861,  & T_{003} &= 0.5619,  & T_{010} &= 0.5555,   & T_{011} &= 0.6663,  & T_{013} &= 0.2585, \\
T_{022} &= -0.4382,  & T_{030} &= 0.3549,  & T_{031} &= -0.0369, & T_{033} &= 0.6534,   & T_{100} &= 0.1546,  & T_{101} &= 0.0528, \\
T_{103} &= 0.1546,   & T_{110} &= 0.2012,  & T_{111} &= 0.1999,  & T_{113} &= 0.2012,   & T_{122} &= -0.1999, & T_{130} &= 0.1546, \\
T_{131} &= 0.0528,   & T_{133} &= 0.1546,  & T_{202} &= -0.0528, & T_{212} &= -0.1999,  & T_{220} &= -0.2012, & T_{221} &= -0.1999, \\
T_{223} &= -0.2012,  & T_{232} &= -0.0528, & T_{300} &= -0.8138, & T_{301} &= -0.5861,  & T_{303} &= -0.3757, & T_{310} &= -0.5555, \\
T_{311} &= -0.6663,  & T_{313} &= -0.2585, & T_{322} &= 0.4382,  & T_{330} &= -0.1688,  & T_{331} &= 0.0369,  & T_{333} &= -0.4672.
\end{aligned}
\end{equation}

The representative states were selected with the aim of ensuring that their purity -- which influences the range of variability of the parameters defining the employed correlation vectors -- allows for an optimal representation of the evolution. Other initial states in their most general form were also examined; all cases confirm the observations made.

\end{document}